\documentclass[format=acmsmall, review=false, screen=true]{acmart}

\usepackage{booktabs} 

\usepackage[ruled]{algorithm2e} 
\renewcommand{\algorithmcfname}{ALGORITHM}
\SetAlFnt{\small}
\SetAlCapFnt{\small}
\SetAlCapNameFnt{\small}
\SetAlCapHSkip{0pt}
\IncMargin{-\parindent}

\acmJournal{JETC}
\acmVolume{9}
\acmNumber{4}
\acmArticle{39}
\acmYear{2010}
\acmMonth{3}
\copyrightyear{2009}

\acmDOI{0000001.0000001}

\received{February 2007}
\received[revised]{March 2009}
\received[accepted]{June 2009}

\usepackage{booktabs}
\usepackage{subcaption}
\usepackage{caption}
\usepackage{multirow}
\usepackage{amsmath}
\usepackage{lscape}
\usepackage{mathtools}
\usepackage{hyperref} 
\usepackage{amsfonts}
\usepackage{graphicx}

\usepackage{multirow}
\usepackage{rotating}
\input{Qcircuit.tex}

\begin{document}
\title{T-count and Qubit Optimized Quantum Circuit Design of the Non-Restoring Square Root Algorithm}  

\author{Edgard Mu\~{n}oz-Coreas and Himanshu Thapliyal}
\affiliation{%
  \institution{University of Kentucky}
  \department{Department of Electrical and Computer Engineering}
  \city{Lexington}
  \state{KY}
  \postcode{40506}
  \country{USA}}
\email{hthapliyal@uky.edu}

\begin{abstract}
Quantum circuits for basic mathematical functions such as the square root are required to implement scientific computing algorithms on quantum computers.  Quantum circuits that are based on Clifford+T gates can easily be made fault tolerant but the T gate is very costly to implement.  As a result, reducing T-count has become an important optimization goal.  Further, quantum circuits with many qubits are difficult to realize, making designs that save qubits and produce no garbage outputs desirable.  In this work, we present a T-count optimized quantum square root circuit with only $2 \cdot n +1$ qubits and no garbage output.  To make a fair comparison against existing work, the Bennett's garbage removal scheme is used to remove garbage output from existing works.  We determined that our proposed design achieves an average T-count savings of $43.44 \%$, $98.95 \%$, $41.06 \%$ and $20.28 \%$ as well as qubit savings of $85.46 \%$, $95.16 \%$, $90.59 \%$ and $86.77 \%$ compared to existing works.   
\end{abstract}


%
%

%
%


\maketitle



\section{Introduction}

Among the emerging computing paradigms, quantum computing appears promising due to its applications in number theory, encryption, search and scientific computation \cite{Cheung} \cite{Bowregard} \cite{Montanaro} \cite{Shparlinski} \cite{Proos} \cite{Seroussi} \cite{Guodong2014quadraticmotivation} \cite{Vandam2000quadraticmotivation} \cite{Hallgren2007quadraticmotivation}.  Quantum circuits for arithmetic operations such as addition, multiplication, square root and fractional powers are required in the quantum circuit implementations of many quantum algorithms  \cite{Quipper} \cite{LIQUi} \cite{Bowregard} \cite{bhaskar} \cite{Cheung}.  For example, arithmetic circuits for the square root can be used in the circuit implementation of quantum algorithms such as those for computing roots of polynomials, evaluating quadratic congruence and the principal ideal problem \cite{Guodong2014quadraticmotivation} \cite{Vandam2000quadraticmotivation} \cite{Hallgren2007quadraticmotivation}.  Quantum square root circuits also reduce the resources needed in the circuit implementations of higher level functions computing the natural logarithm  \cite{bhaskar}.  An efficient quantum circuit of the natural logarithm has use in quantum algorithms such as those for Pell's equation and the principal ideal problem \cite{Hallgren2007quadraticmotivation}.  The design of quantum circuits for arithmetic operations such as addition and multiplication have received notable attention in the literature.  However, the design of quantum circuits for crucial arithmetic functions such as the square root is still at an initial stage.

Reliable quantum circuits must be able to tolerate noise errors \cite{Webster} \cite{Zhou-T} \cite{Paler_DAC} \cite{Polian_DAC}.   Fault tolerant quantum gates (such as Clifford+T gates) and quantum error correcting codes can be used to make quantum circuits resistant to noise errors \cite{Mosca2} \cite{Miller} \cite{Maslov} \cite{Cody} \cite{PalerIOP} \cite{Devitt}.  However, the increased tolerance to noise errors comes with the increased implementation overhead associated with the quantum T gate \cite{Mosca2} \cite{PalerIOP} \cite{Zhou-T} \cite{Devitt}.  Because of the increased cost to realize the T gate, T-count has become an important performance measure for fault tolerant quantum circuit design \cite{Mosca2} \cite{Gosset}. Further, existing quantum computers have few qubits and large-scale quantum computers are difficult to realize \cite{IBM_quantum} \cite{Monroe}.  As a result, the total number of qubits required by a quantum circuit is an important performance measure.  Quantum circuits have overhead called ancillae and garbage output that add to the total number of qubits of a quantum circuit.  Any constant inputs in the quantum circuit are called ancillae.  Garbage output exist in the quantum circuit to preserve one-to-one mapping.  Garbage output are not primary inputs or useful outputs.  Minimizing the overhead from ancillae and garbage output is a means to reduce overall qubit cost of a quantum circuit.      


The design of quantum circuits for the calculation of the square root has only recently begun to be addressed in the literature.  A design for the calculation of the square root based on the Newton approximation algorithm is presented in \cite{bhaskar}.  While an interesting design, the implementation requires $5 \cdot \lceil log_2(b) \rceil$ multiplications and $3 \cdot \lceil log_2(b) \rceil$ additions (where $b$ is the number of bits of accuracy in the solution and $b \geq 4$) \cite{bhaskar}.  This arithmetic operation cost translates into significant T gate and qubit cost.  The design in \cite{Sultana} presents a quantum circuit for calculating the square root based on the non-restoring square root algorithm.  This design requires only $\frac{n}{2}$ additions or subtractions making the design far more efficient than the design in \cite{bhaskar} in terms of qubits and T gates.  However, the design in \cite{Sultana} does not include the additional ancillae and T gate costs required for removing garbage outputs.  Additional recent quantum circuit designs for calculating the square root presented in \cite{ANANTHALAKSHM} also require only $\frac{n}{2}$ controlled subtraction operations.  The designs in \cite{ANANTHALAKSHM} are based on the non-restoring square root algorithm.  Thus, the designs presented in \cite{ANANTHALAKSHM} also offer more efficient alternatives to the design in \cite{bhaskar} in terms of qubits and T gates.  One of the designs presented in \cite{ANANTHALAKSHM} has been optimized for gate count further reducing its T-gate cost.  Both designs presented in \cite{ANANTHALAKSHM} produce significant garbage output.   In \cite{ANANTHALAKSHM} for both designs the qubit and T gate cost associated with removing this garbage output was not considered in the circuit cost calculations.  
Thus, the quantum square root circuits in \cite{Sultana} and \cite{ANANTHALAKSHM}  have significant overhead in terms of T-count and qubits.  \textit{To overcome the limitations of existing designs, we present the design of a quantum square root circuit that is garbageless, requires $2 \cdot n + 1$ qubits and is optimized for T-count.  The quantum square root circuit based on our proposed design is compared and is shown to be better than the existing designs of quantum square root circuit in terms of both T-count and qubits.}

This paper is organized as follows:  Section \ref{sqrt-ref} presents background information on the Clifford+T gates.  In Section \ref{sqrt-builder} we present the design of our proposed quantum square root circuit.  In Section \ref{sqrt-cost} our proposed design is compared to the existing work.

\section{Background}
\label{sqrt-ref}

\begin{table}[tbhp]
\centering
\caption{The Clifford + T gates}
\label{Clifford table}
{ \small
\begin{tabular}{|c|c|c|}
\hline

\textbf{ Type of Gate} &\textbf{Symbol} & \textbf{Matrix}\\
 \hline
 NOT gate  & $X$   &    $\begin{bmatrix}
    0 & 1  \\
    1 & 0 
  \end{bmatrix}$ \\
 
 \hline
 Hadamard gate   & $H$   &   $ \frac{1}{\sqrt{2}}
  \begin{bmatrix}
    1 & 1  \\
    1 & -1 
  \end{bmatrix} $\\
  
\hline
$T$ gate  & $T$   &  $\begin{bmatrix}
    1 & 0  \\
    0 & e^{i.\frac{\pi}{4}} 
  \end{bmatrix} $\\ 
 
\hline
$T$ gate Hermitian transpose  & $T^{\dag}$   &  $\begin{bmatrix}
    1 & 0  \\
    0 & e^{-i.\frac{\pi}{4}} 
  \end{bmatrix} $\\  
 
 \hline
 Phase gate & $S$   &  $  \begin{bmatrix}
    1 & 0  \\
    0 & i 
  \end{bmatrix}$\\
 \hline

 Phase gate Hermitian transpose  & $S^{\dag}$   &  $  \begin{bmatrix}
    1 & 0  \\
    0 & -i 
  \end{bmatrix}$\\
 \hline
 
CNOT gate  & $C$    & $\begin{bmatrix}
    1 & 0 & 0 & 0 \\
    0 & 1 & 0 & 0\\
    0 & 0 & 0 & 1 \\
    0 & 0 & 1 & 0
  \end{bmatrix} $ \\
 
 \hline
\end{tabular}
}

\end{table}

\begin{figure}[tbhp]

\includegraphics[width=3.5in]{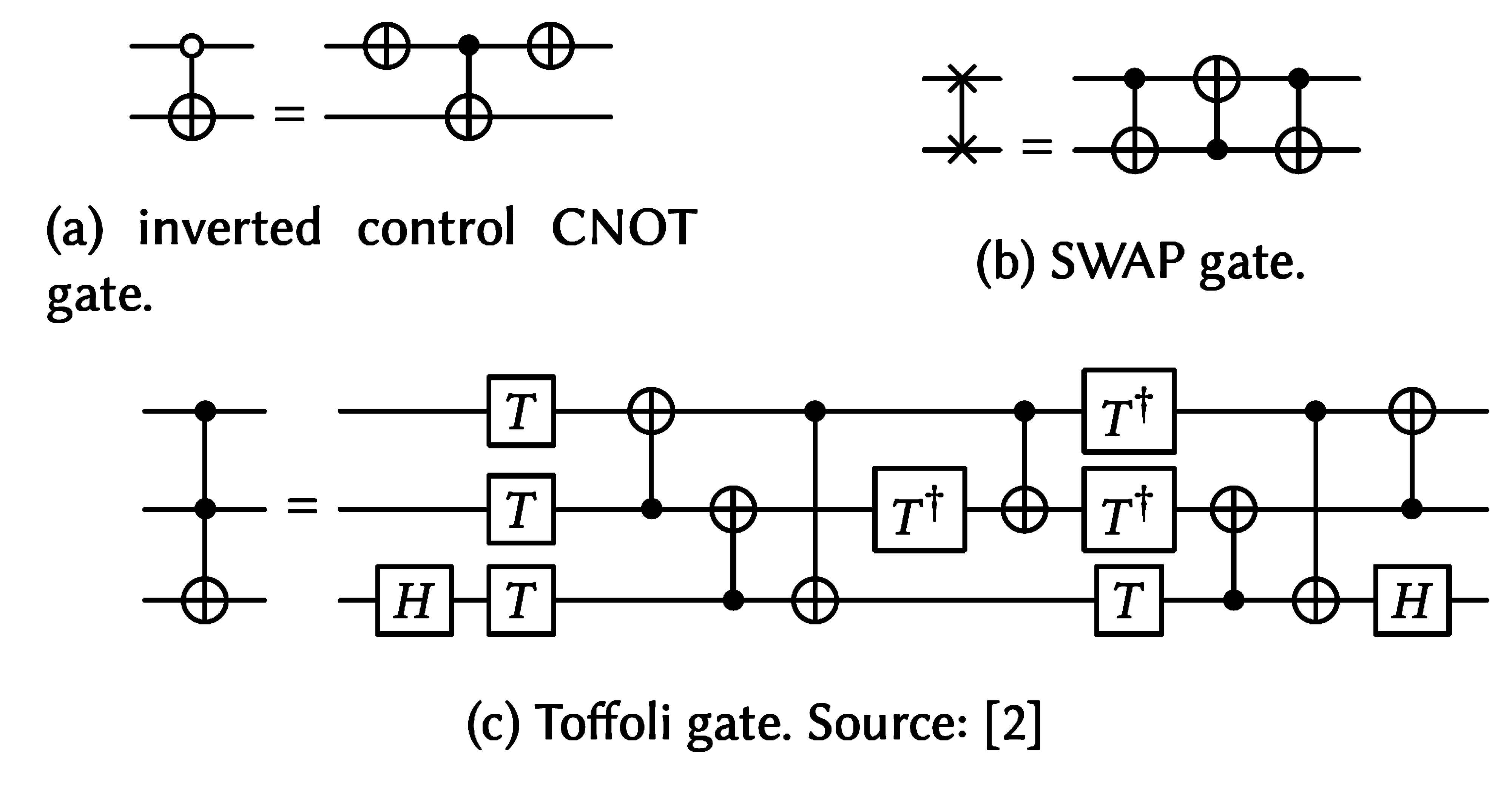}

\caption{The fault tolerant Clifford + T implementations of quantum logic gates used in this work. The Toffoli gate shown has a T-count of 7 and a T-depth of 3.}
\label{sqrt-FIG1}
\end{figure}

\subsection{Fault Tolerant Quantum Circuits}

The fault tolerant Clifford+T gate set is used in fault tolerant quantum circuit design.  Table \ref{sqrt-FIG1} shows the gates that make up the Clifford+T gate set.  \textit{The quantum square root circuit proposed in this work is composed of the quantum NOT gate, Feynman (CNOT) gate, inverted control CNOT gate, SWAP gate and Toffoli gate.}  Table \ref{Clifford table} illustrates that the CNOT gate and NOT gate are Clifford+T gates.  The Clifford+T implementation of the inverted control CNOT gate, SWAP gate and Toffoli gate are shown in Figure \ref{sqrt-FIG1}.  In this work, we use the Clifford+T implementation of the Toffoli gate presented in \cite{Maslov}.  The inverted control CNOT gate and the SWAP gate are 2 input, 2 output logic gates that have the mapping $A,B$ to $A, \overline{A} \oplus B$ and $A,B$ to $B,A$ respectively.  The Toffoli gate is a 3 input, 3 output logic gate and has the mapping $A,B,C$ to $A,B,A \cdot B \oplus C$.

Fault tolerant quantum circuit performance is evaluated in terms of T-count and T-depth because the implementation costs of the T gate is significantly greater than the implementation costs of the other Clifford+T gates \cite{Mosca2} \cite{PalerIOP} \cite{Zhou-T} \cite{Devitt} \cite{Gosset}.  \textit{T-count is the total number of T gates or Hermitian transposes of the T gate in a quantum circuit.}  As illustrated in Figure \ref{sqrt-FIG1}, the inverted control CNOT gate and the SWAP gate both have a T-count of $0$ while the Toffoli gate has a T-count of $7$.  \textit{T-depth is the number of T gate layers in the circuit, where a layer consists of quantum operations that can be performed simultaneously.}  As illustrated in Figure \ref{sqrt-FIG1}, the inverted control CNOT gate and the SWAP gate both have a T-depth of $0$.  The Toffoli gate has a T-depth of $3$ because the most T gate layers encountered by any qubit in the Toffoli gate is $3$. 

\section{Design of the Proposed Quantum Square Root Circuit}
\label{sqrt-builder}

The proposed quantum square root circuit calculates the square root by implementing the non-restoring square root algorithm.  The non-restoring square root algorithm is illustrated in Figure \ref{sqrt-table:70} (Algorithm 1).  Researchers have demonstrated the correctness of the non-restoring square root algorithm through functionally correct circuit implementations such as those in \cite{Sultana} \cite{ANANTHALAKSHM} and \cite{SAMAVI2008sqrt}.  A specific example illustrating how Algorithm 1 calculates the square root of a number $a$ is available in Appendix \ref{sqrt-explain}.

\begin{figure}[hbtp]
\small
\begin{tabular}{ll}
\\ \midrule
\multicolumn{2}{l}{\textbf{Algorithm 1:} Non-restoring square root algorithm} \\ \toprule
\multicolumn{2}{l}{\textbf{Requirements:} $a$ must be a positive binary value in $2's$ complement that has an even bit length $n$} \\
\multicolumn{2}{l}{\textbf{Input:} $a$.  $a$ is incrementally loaded into $R$ starting from the most significant bit.} \\
\multicolumn{2}{l}{\textbf{Outputs:} $\sqrt{a}$ and the remainder from calculating $\sqrt{a}$.  The $\sqrt{a}$ is an $\frac{n}{2}$ bit value in $F$. }\\
\multicolumn{2}{l}{We will use the variable $Y$ to represent the $\sqrt{a}$.  $R$ will have the $n$ bit remainder.} \\

1 & \textbf{Function} Non-Restoring($a$) \\ 
2 & \qquad $R = 0^{n-2} a_{n-1}a_{n-2}$ //where $0^{n-2}$ are $n-2$ zeros. $a_{n-1}$ is the most significant bit of $a$.\\
3 & \qquad $F = 0^{n-2} 01$ //where $0^{n-2}$ are $n-2$ zeros. \\
4 & \qquad $R = R - F$ \\ 
5 & \\
6 & \qquad \textbf{For} $i = \frac{n}{2} -1 \text{ to } 1$ \\ 
7 & \qquad \qquad \textbf{If}$(R < 0)$ \\
8 & \qquad \qquad \qquad $Y_{i} = 0$ \\  
9 & \qquad \qquad \qquad $R =  0^{2 \cdot i - 2} R_{n-1-2 \cdot i} \cdots R_0 a_{2 \cdot i-1} a_{2 \cdot i -2} $ //where $0^{2 \cdot i - 2}$ are $2 \cdot i - 2$ zeros. \\
10 & \qquad \qquad \qquad \qquad //Values $R_{n-1-2 \cdot i}$ through $R_0$ of $R$ are shifted and reused.   \\
11 & \qquad \qquad \qquad $F = 0^{i+ \frac{n}{2} - 2} Y_{\frac{n}{2} -1} Y_{\frac{n}{2} -2} \cdots Y_{i+1} Y_i 11$ //where $0^{i+ \frac{n}{2} - 2}$ are $i+ \frac{n}{2} - 2$ zeros.  \\
12 & \qquad \qquad \qquad \qquad //$Y_{\frac{n}{2}-1}$ is the most significant bit of $Y$.   \\
13 & \qquad \qquad \qquad $R = R + F$ \\
14 & \qquad \qquad \textbf{Else} \\
15 & \qquad \qquad \qquad $Y_{i} = 1$ \\
16 & \qquad \qquad \qquad $R = 0^{2 \cdot i - 2} R_{n-1-2 \cdot i} \cdots R_0 a_{2 \cdot i-1} a_{2 \cdot i -2} $ //where $0^{2 \cdot i - 2}$ are $2 \cdot i - 2$ zeros. \\
17 & \qquad \qquad \qquad \qquad //Values $R_{n-1-2 \cdot i}$ through $R_0$ of $R$ are shifted and reused.   \\
18 & \qquad \qquad \qquad $F = 0^{i+ \frac{n}{2} - 2} Y_{\frac{n}{2} -1} Y_{\frac{n}{2} -2} \cdots Y_{i+1} Y_i 01$ //where $0^{i+ \frac{n}{2} - 2}$ are $i+ \frac{n}{2} - 2$ zeros. \\
19 & \qquad \qquad \qquad \qquad //$Y_{\frac{n}{2}-1}$ is the most significant bit of $Y$.   \\
20 & \qquad \qquad \qquad $R = R - F$ \\
21 & \qquad \qquad \textbf{End} \\
22 & \qquad \textbf{End} \\ 
23 & \\
24 & \qquad \textbf{If}$(R < 0)$ \\
25 & \qquad \qquad $Y_0 = 0$ \\
26 & \qquad \qquad $F = 0^{\frac{n}{2} - 2} Y_{\frac{n}{2} -1} Y_{\frac{n}{2} -2} \cdots Y_1 Y_0 01$ //where $0^{\frac{n}{2} - 2}$ are $\frac{n}{2} - 2$ zeros. \\
27 & \qquad \qquad \qquad //$Y_{\frac{n}{2}-1}$ is the most significant bit of $Y$.   \\
28 & \qquad \qquad $R = R + F$ \\
29 & \qquad \textbf{Else} \\
30 & \qquad \qquad $Y_0 = 1$ \\
31 & \qquad \qquad $F = 0^{\frac{n}{2} - 2} Y_{\frac{n}{2} -1} Y_{\frac{n}{2} -2} \cdots Y_1 Y_0 01$ //where $0^{\frac{n}{2} - 2}$ are $\frac{n}{2} - 2$ zeros.  \\
32 & \qquad \qquad \qquad //$Y_{\frac{n}{2}-1}$ is the most significant bit of $Y$.   \\
33 & \qquad \textbf{End} \\ 
34 & \textbf{Return:} $R,F$; \\ \bottomrule

\end{tabular}

\caption{The non-restoring square root algorithm.  The algorithm has been adapted from the presentation shown in \cite{SAMAVI2008sqrt}.}
\label{sqrt-table:70}
\end{figure}

We now present the design of our proposed quantum square root circuit.  The proposed method is garbageless and requires fewer qubits than existing designs.  The proposed circuit also has a lower T-count compared to existing designs.  Consider the square root of the number $a$.  We represent $a$ as a positive binary value in $2's$ complement that has an even bit length $n$.  $a$ is stored in quantum register $\ket{R}$.  Further, let $\ket{F}$ be a quantum register of size $n$ initialized to $1$ and let $\ket{z}$ be a 1 qubit ancillae set to $0$.  At the end of computation, quantum register locations $\ket{F_{\frac{n}{2}+1}}$ through $\ket{F_2}$ of $\ket{F}$ will have the value $Y$ ($\sqrt{a}$).  In addition, quantum register $\ket{R}$ that initially stored $a$ will have the remainder from the calculation of $\sqrt{a}$.  Lastly, quantum register $\ket{z}$ and the remaining register locations $\ket{F_{n-1}}$ through $\ket{F_{\frac{n}{2} + 2}}$ and $\ket{F_{1}}$ through $\ket{F_{0}}$ of $\ket{F}$ are restored to their initial values at the end of computation.    

The proposed methodology is generic and can design a quantum square root circuit of any size.  The steps involved in the proposed methodology are presented for finding the square root of the value $a$.  \textit{The proposed quantum square root circuit is divided into three parts: (i) Part 1: Initial Subtraction, (ii) Part 2: Conditional Addition or Subtraction, and (iii) Part 3: Remainder Restoration.}  A quantum circuit is generated for each part of the design.  Each part implements the following portions of the non-restoring square root algorithm shown in Algorithm 1.  

\begin{itemize}

\item \textbf{Part 1: Initial Subtraction} This part executes the statements before the FOR loop in Algorithm 1.  Also,  this part executes the first iteration of the FOR loop in Algorithm 1. 

\item \textbf{Part 2: Conditional Addition or Subtraction} This part executes the remaining $\frac{n}{2}-2$ iterations of the FOR loop in Algorithm 1.  Thus, Part 2 will be iterated a total of $\frac{n}{2}-2$ times.

\item \textbf{Part 3: Remainder Restoration} This part implements the IF statement that follows the FOR loop in Algorithm 1.

\end{itemize}  

Figure \ref{sqrt-FIG7} shows a generic example of how the parts of our quantum square root circuit are combined to implement Algorithm 1.  The detailed quantum circuit designs of Part 1, Part 2, and Part 3 required to implement Algorithm 1 are explained as follows:

\begin{figure}[htbp]

\includegraphics[width=3in]{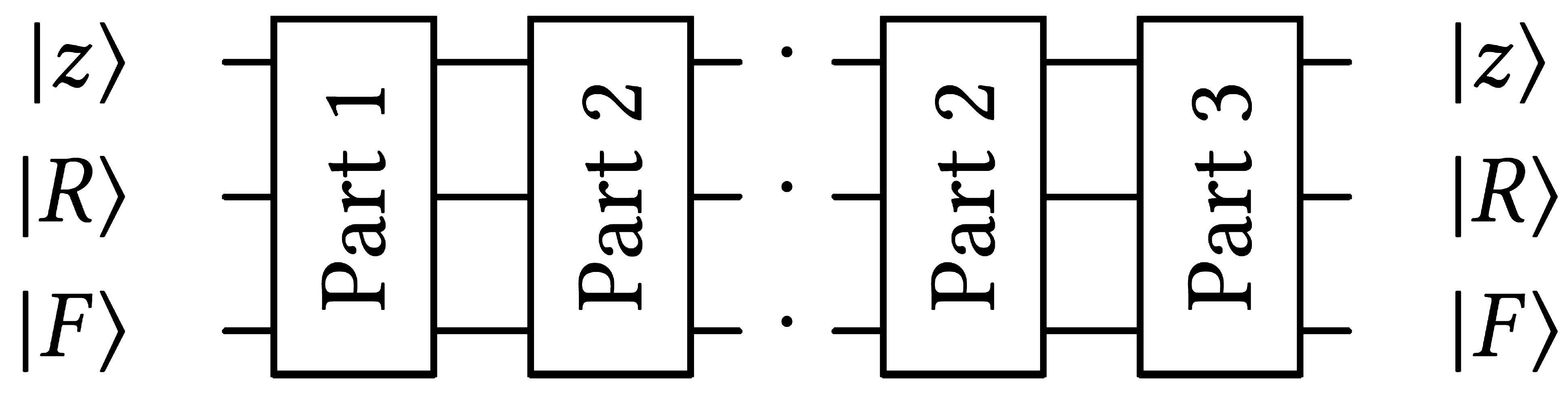}

\caption{Example of the complete proposed square root circuit. }
\label{sqrt-FIG7}
\end{figure}

\subsection{Part 1: Initial Subtraction}

\begin{figure}[t!bhp]

 \begin{subfigure}[th]{2in}
\centering
\includegraphics[width=2.25in]{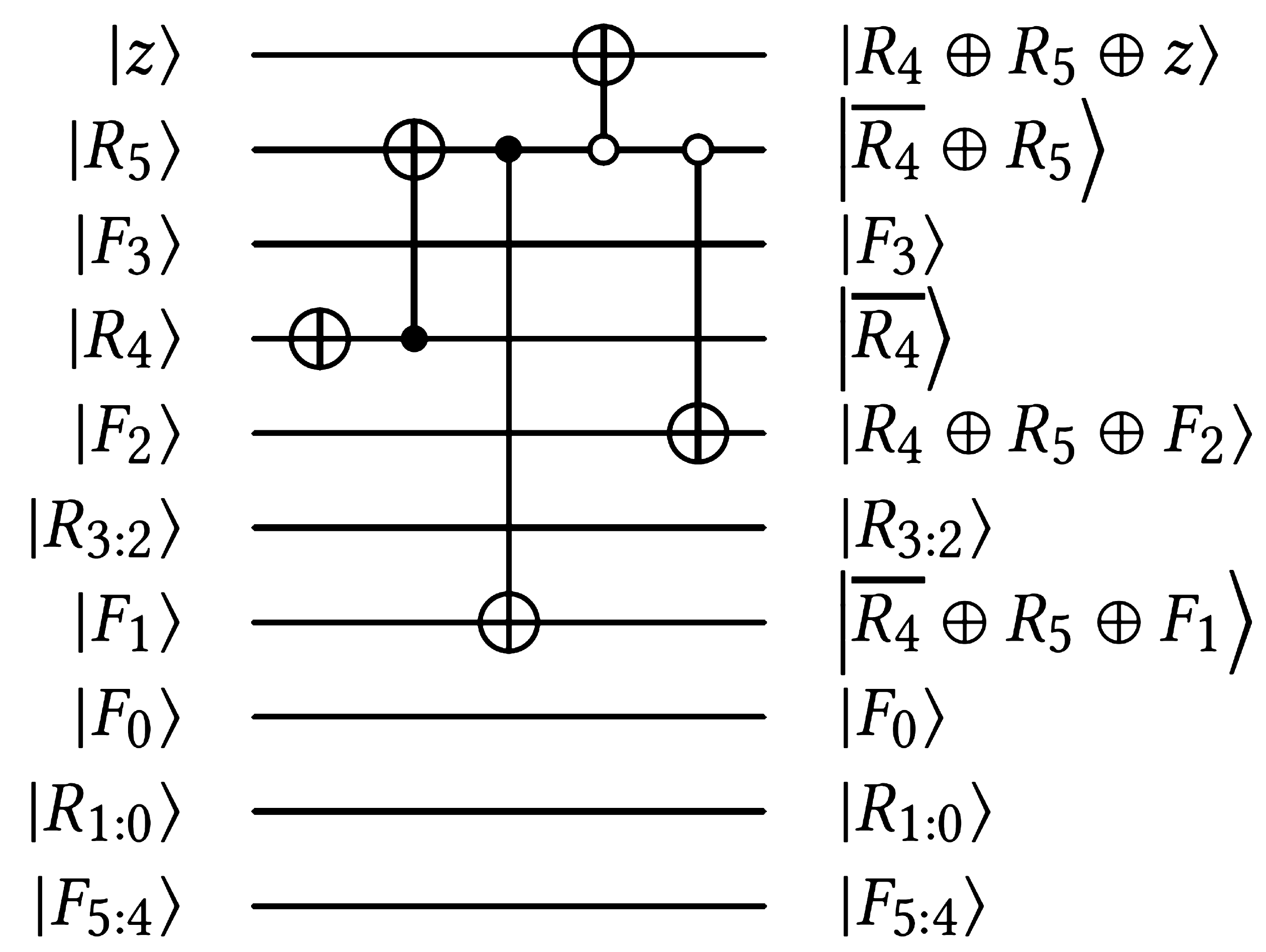}

\caption{After Steps 1 through 5}
\end{subfigure} 
\\ 
\begin{subfigure}[th]{4in}
\flushleft
\includegraphics[width=3.55in]{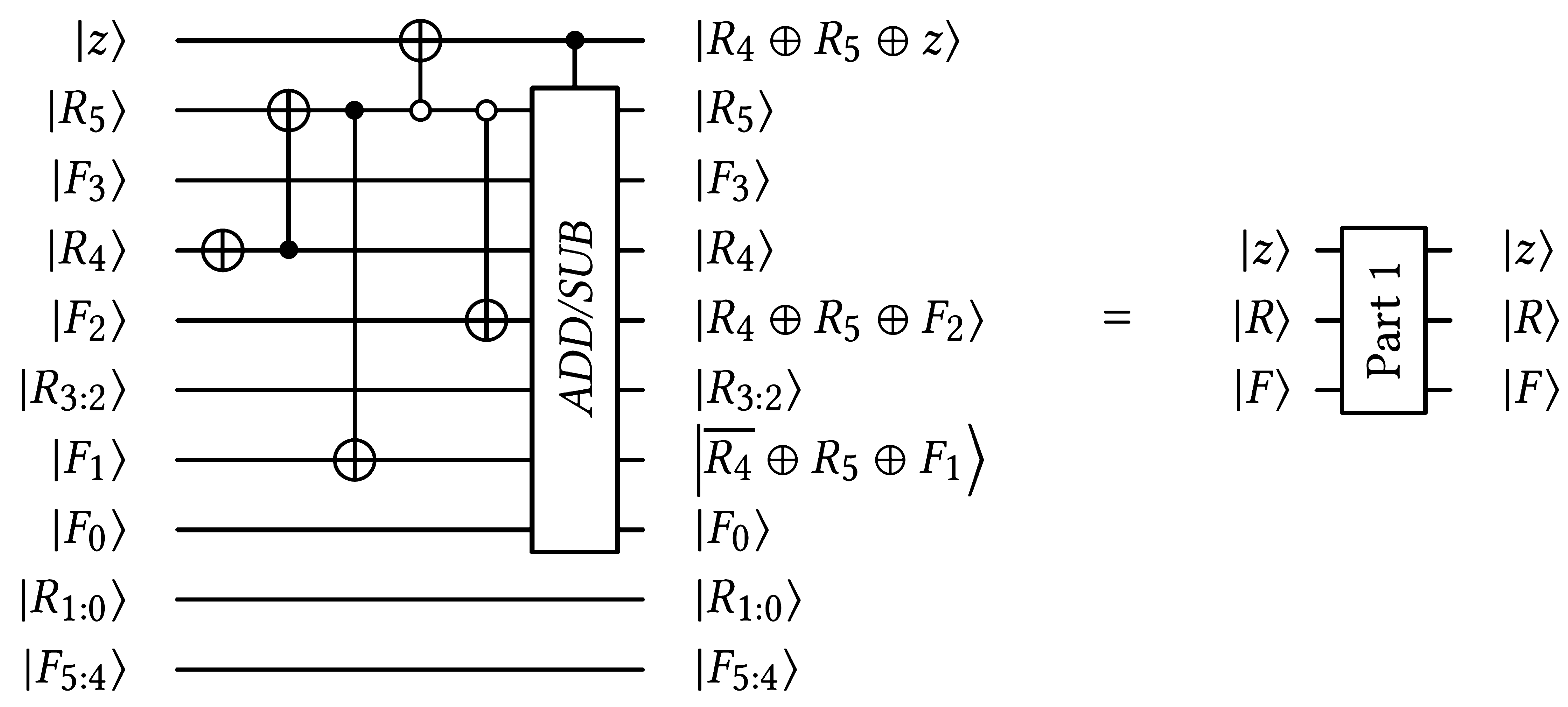}

\caption{After Step 6.  Quantum circuit and graphical representation are shown.}
\end{subfigure}
\caption{Circuit generation of Part 1 of the proposed quantum square root circuit: Steps 1-6. To keep Figures compact quantum register locations (such as $\ket{F_5}$ through $\ket{F_4}$) are represented as a single line and labeled accordingly (such as with $\ket{F_{5:4}}$) where possible.}
\label{sqrt-FIG2}
\end{figure}


This part only occurs once.   The quantum circuit for Part 1 takes quantum registers $\ket{R}$, $\ket{F}$ and $\ket{z}$ as inputs.   Part 1 has six steps.  Figure \ref{sqrt-FIG2} illustrates the generation of Part 1 with an example of a 6 bit square root circuit.   

\begin{itemize}

\item Step 1:  At location $\ket{R_{n-2}}$ apply a quantum NOT gate.  

\item Step 2:  At locations $\ket{R_{n-1}}$ and $\ket{R_{n-2}}$  apply a CNOT gate such that the location $\ket{R_{n-2}}$ is unchanged while location $\ket{R_{n-1}}$ now has the value $\ket{\overline{R_{n-2}} \oplus R_{n-1}}$ (where $\overline{R_{n-2}} \oplus R_{n-1} \equiv \overline{Y_{\frac{n}{2}-1}}$).  Step 1 and Step 2 implement lines 2 through 4 of Algorithm 1.  

\item Step 3: At locations $\ket{R_{n-1}}$ and $\ket{F_1}$ apply a CNOT gate such that the location $\ket{R_{n-1}}$ is unchanged while location $\ket{F_1}$ now has the value $\ket{\overline{R_{n-2}} \oplus R_{n-1} \oplus F_1}$. If $\ket{R_{n-1}} = 1$, this step partially implements line 11 of Algorithm 1 because the value at location $\ket{F_1}$ $ \left( \ket{\overline{R_{n-2}} \oplus R_{n-1} \oplus F_1} \right)$ simplifies to $\ket{1}$.  Otherwise this step helps to implement line 18 of Algorithm 1 because the value at location $\ket{F_1}$ $\left( \ket{\overline{R_{n-2}} \oplus R_{n-1} \oplus F_1} \right)$ simplifies to $\ket{0}$ when $\ket{R_{n-1}} = 0$.


\item Step 4: At locations $\ket{R_{n-1}}$ and $\ket{z}$ apply an inverted control CNOT gate such that the location $\ket{R_{n-1}}$ is unchanged while location $\ket{z}$ now has the value $\ket{\overline{\overline{R_{n-2}} \oplus R_{n-1} \oplus z}}$ which simplifies to $\ket{R_{n-2} \oplus R_{n-1} \oplus z}$ (where $R_{n-2} \oplus R_{n-1} \oplus z \equiv Y_{\frac{n}{2}-1}$).  This step prepares register $\ket{z}$ for use in subsequent steps.

\item Step 5: At locations $\ket{R_{n-1}}$ and $\ket{F_2}$ apply an inverted control CNOT gate such that the location $\ket{R_{n-1}}$ is unchanged while location $\ket{F_2}$ now has the value $\ket{\overline{\overline{R_{n-2}} \oplus R_{n-1}} \oplus F_2}$ which simplifies to $\ket{R_{n-2} \oplus R_{n-1} \oplus F_2}$ (where $R_{n-2} \oplus R_{n-1} \oplus F_2 \equiv Y_{\frac{n}{2}-1}$).  If $\ket{R_{n-1}} = 1$ this step completes execution of line 11 of Algorithm 1 and quantum register $\ket{F}$ will have the value: $\ket{0} \cdots \ket{0}  \ket{Y_{\frac{n}{2}-1}} \ket{1} \ket{1}$.  Conversely, if  $\ket{R_{n-1}} = 0$ this step completes execution of line 18 of Algorithm 1 and quantum register $\ket{F}$ will have the value: $\ket{0} \cdots \ket{0}  \ket{Y_{\frac{n}{2}-1}} \ket{0} \ket{1}$.

\item Step 6: this step has two sub-steps. 

\begin{itemize}

\item Step 1: At locations $\ket{R_{n-1}}$ through $\ket{R_{n-4}}$ of register $\ket{R}$ and locations $\ket{F_{3}}$ through $\ket{F_{0}}$ of register  $\ket{F}$ apply the quantum conditional addition or subtraction (\textit{ADD/SUB}) circuit such that locations $\ket{F_{3}}$ through $\ket{F_{0}}$ are unchanged while locations $\ket{R_{n-1}}$ through $\ket{R_{n-4}}$ will hold the results of computation.  

\item Step 2: At location $\ket{z}$ apply the quantum \textit{ADD/SUB} circuit such that the operation of the circuit is conditioned on the value at location $\ket{z}$.  Location $\ket{z}$ is unchanged.

\end{itemize}

After this step, if $\ket{R_{n-1}} = 1$, the quantum register $\ket{R}$ will equal $\ket{R} + \ket{F}$ (line 13 of Algorithm 1).  If $\ket{R_{n-1}} = 0$, the quantum register $\ket{R}$ will equal $\ket{R} - \ket{F}$ (line 20 of Algorithm 1). 


\end{itemize}

\subsection{Part 2: Conditional Addition or Subtraction}

\begin{figure}[tbhp]
\centering
  \begin{subfigure}[th]{2.1in}

\includegraphics[width=2.25in]{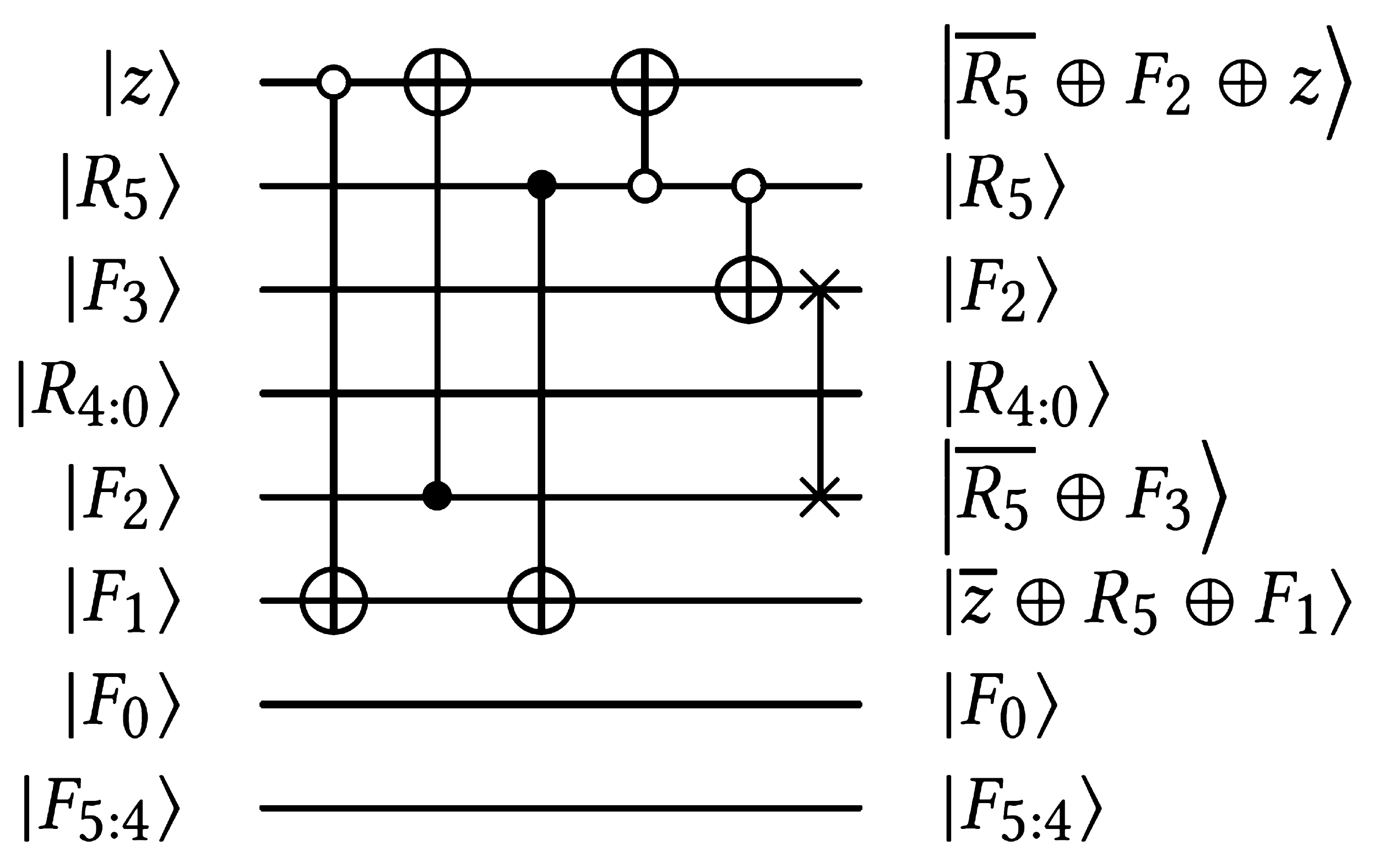}

\caption{After Steps 1 through 6}
\end{subfigure}
\\
\begin{subfigure}[th]{4in}

\includegraphics[width=3.55in]{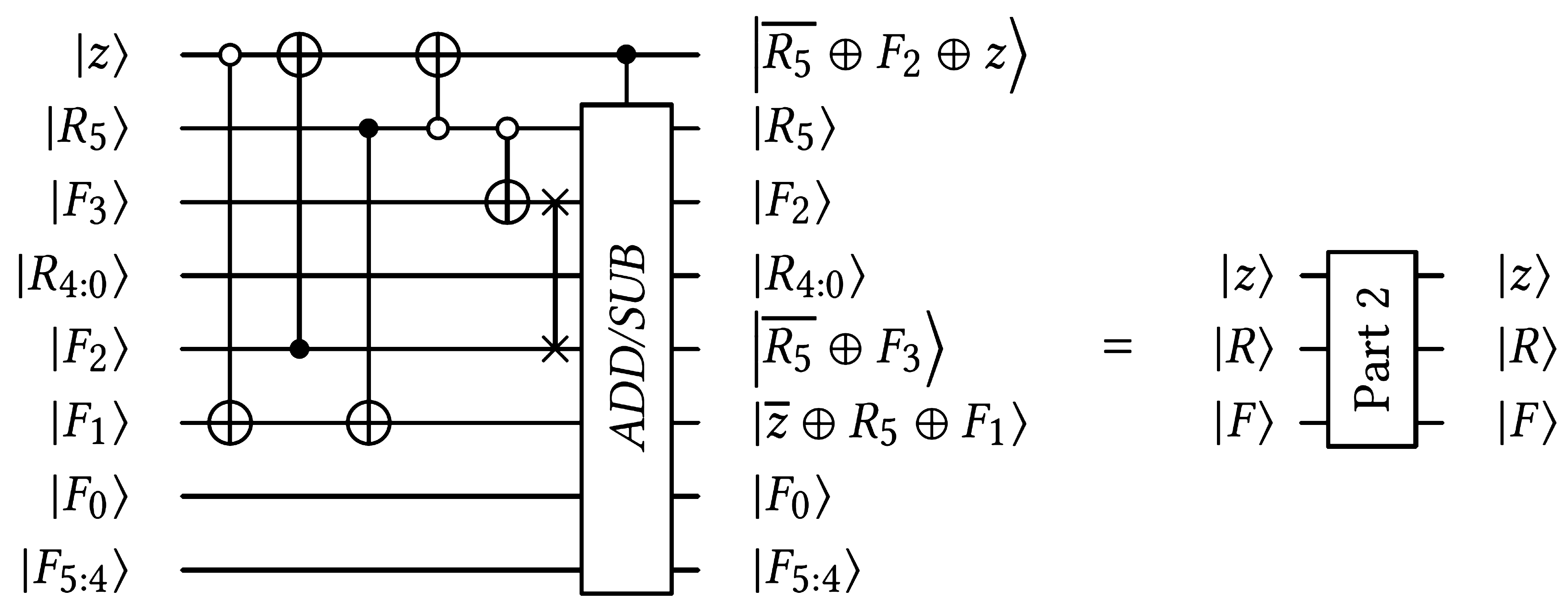}

\caption{After Step 7.  Quantum circuit and graphical representation are shown.}
\end{subfigure}

\caption{Circuit generation of Part 2 of the proposed quantum square root circuit: Steps 1-7. To keep Figures compact quantum register locations (such as $\ket{R_4}$ through $\ket{R_0}$) are represented as a single line and labeled accordingly (such as with $\ket{R_{4:0}}$) where possible.}
\label{sqrt-FIG3}
\end{figure}

This part is repeated a total of $\frac{n}{2}-2$ times.   The quantum circuit for each iteration of Part 2 takes quantum registers $\ket{R}$, $\ket{F}$ and $\ket{z}$ as inputs.  Part 2 has seven steps.  Figure \ref{sqrt-FIG3} illustrates the generation of Part 2 with an example of a 6 bit square root circuit.  We show the steps for iteration $i$ where $2 \leq i \leq \frac{n}{2}-1$. 

\begin{itemize}

\item Step 1: At locations $\ket{z}$ and $\ket{F_1}$ apply an inverted control CNOT gate such that the location $\ket{z}$ is unchanged while location $\ket{F_1}$ now has the value $\ket{\overline{z} \oplus F_1}$.  This step restores $\ket{F_1}$ to its initial value such that $\ket{F}$ has the value: $ \ket{0} \cdots \ket{0} \ket{Y_{\frac{n}{2}-1}} \cdots \ket{Y_{\frac{n}{2}-i+1}}  \ket{0} \ket{1}$. 

\item Step 2: At locations $\ket{F_2}$ and $\ket{z}$ apply a CNOT gate such that the location $\ket{F_2}$ is unchanged while location $\ket{z}$ now has the value $\ket{F_2 \oplus z}$ which reduces to $0$. Steps 1 and 2 prepare $\ket{z}$ and $\ket{F}$ for iteration $i$ of the FOR loop in Algorithm 1.  

\item Step 3: At locations $\ket{R_{n-1}}$ and $\ket{F_1}$ apply a CNOT gate such that the location $\ket{R_{n-1}}$ is unchanged while location $\ket{F_1}$ now has the value $\ket{ \overline{z} \oplus R_{n-1} \oplus F_1}$.  If $\ket{R_{n-1}} = 1$, this step partially implements line 11 of Algorithm 1 because the value at location $\ket{F_1}$ $\left(\ket{\overline{z} \oplus R_{n-1} \oplus F_1} \right)$ simplifies to $\ket{1}$.  Otherwise this step helps to implement line 18 of Algorithm 1 because the value at location $\ket{F_1}$ $\left( \ket{\overline{z} \oplus R_{n-1} \oplus F_1} \right)$ simplifies to $\ket{0}$ when $\ket{R_{n-1}} = 0$.

\item Step 4: At locations $\ket{R_{n-1}}$ and $\ket{z}$ apply an inverted control CNOT gate such that the location $\ket{R_{n-1}}$ is unchanged while location $\ket{z}$ now has the value $\ket{\overline{R_{n-1}} \oplus F_2 \oplus z}$ (where $\overline{R_{n-1}} \oplus F_2 \oplus z \equiv Y_{\frac{n}{2}-i}$).  This step prepares register $\ket{z}$ for use in subsequent steps.

\item Step 5: At locations $\ket{R_{n-1}}$ and $\ket{F_{i+1}}$ apply an inverted control CNOT gate such that the location $\ket{R_{n-1}}$ is unchanged while location $\ket{F_{i+1}}$ now has the value $\ket{\overline{R_{n-1}} \oplus F_{i+1}}$ (where $\overline{R_{n-1}} \oplus F_{i+1} \equiv Y_{\frac{n}{2}-i}$).  If $\ket{R_{n-1}} = 1$ this step continues execution of line 11 of Algorithm 1 and quantum register $\ket{F}$ will have the value: $\ket{0} \cdots \ket{0}  \ket{Y_{\frac{n}{2}-i}} \ket{Y_{\frac{n}{2}-1}} \cdots \ket{Y_{\frac{n}{2}-i+1}} \ket{1} \ket{1}$.  Conversely, if  $\ket{R_{n-1}} = 0$ this step continues execution of line 18 of Algorithm 1 and quantum register $\ket{F}$ will have the value: $\ket{0} \cdots \ket{0}  \ket{Y_{\frac{n}{2}-i}} \ket{Y_{\frac{n}{2}-1}} \cdots \ket{Y_{\frac{n}{2}-i+1}} \ket{0} \ket{1}$.

\item Step 6: For $j = i+1 \text{ to } 3$:

At locations $\ket{F_{j}}$ and $\ket{F_{j-1}}$ apply a quantum SWAP gate such the values at locations $\ket{F_{j}}$ and $\ket{F_{j-1}}$ switch locations.  This step reorders the positions of the bits of $Y$ in $\ket{F}$.  Now when $\ket{R_{n-1}} = 1$, quantum register $\ket{F}$ will have the value: $\ket{0} \cdots \ket{0}  \ket{Y_{\frac{n}{2}-1}} \cdots \ket{Y_{\frac{n}{2}-i}} \ket{1} \ket{1}$ and when  $\ket{R_{n-1}} = 0$ quantum register $\ket{F}$ will have the value: $\ket{0} \cdots \ket{0}  \ket{Y_{\frac{n}{2}-1}} \cdots \ket{Y_{\frac{n}{2}-i}} \ket{0} \ket{1}$.  This step completes the execution of line 11 of Algorithm 1 or line 18 of Algorithm 1 depending on the value of $\ket{R_{n-1}}$.

\item Step 7: this Step has two sub-steps.  

\begin{itemize}

\item Step 1: At locations $\ket{R_{n-1}}$ through $\ket{R_{n-2 \cdot i-2}}$ of register $\ket{R}$ and $\ket{F_{2 \cdot i+1}}$ through $\ket{F_{0}}$ of register  $\ket{F}$ apply the quantum conditional addition or subtraction (\textit{ADD/SUB}) circuit such that locations $\ket{F_{2 \cdot i+1}}$ through $\ket{F_{0}}$ are unchanged while locations $\ket{R_{n-1}}$ through $\ket{R_{n-2 \cdot i-2}}$ will hold the results of computation.


\item Step 2: At location $\ket{z}$ apply the quantum \textit{ADD/SUB} circuit such that the operation of the circuit is conditioned on the value at location $\ket{z}$.  Location $\ket{z}$ is unchanged.

\end{itemize}

After this step, if $\ket{R_{n-1}} = 1$, the quantum register $\ket{R}$ will equal $\ket{R} + \ket{F}$ (line 13 of Algorithm 1).  If $\ket{R_{n-1}} = 0$, the quantum register $\ket{R}$ will equal $\ket{R} - \ket{F}$ (line 20 of Algorithm 1). 

\end{itemize}

\subsection{Part 3: Remainder Restoration}

\begin{figure}[tbhp]
\centering
\begin{subfigure}[th]{2.1in}

\includegraphics[width=2.15in]{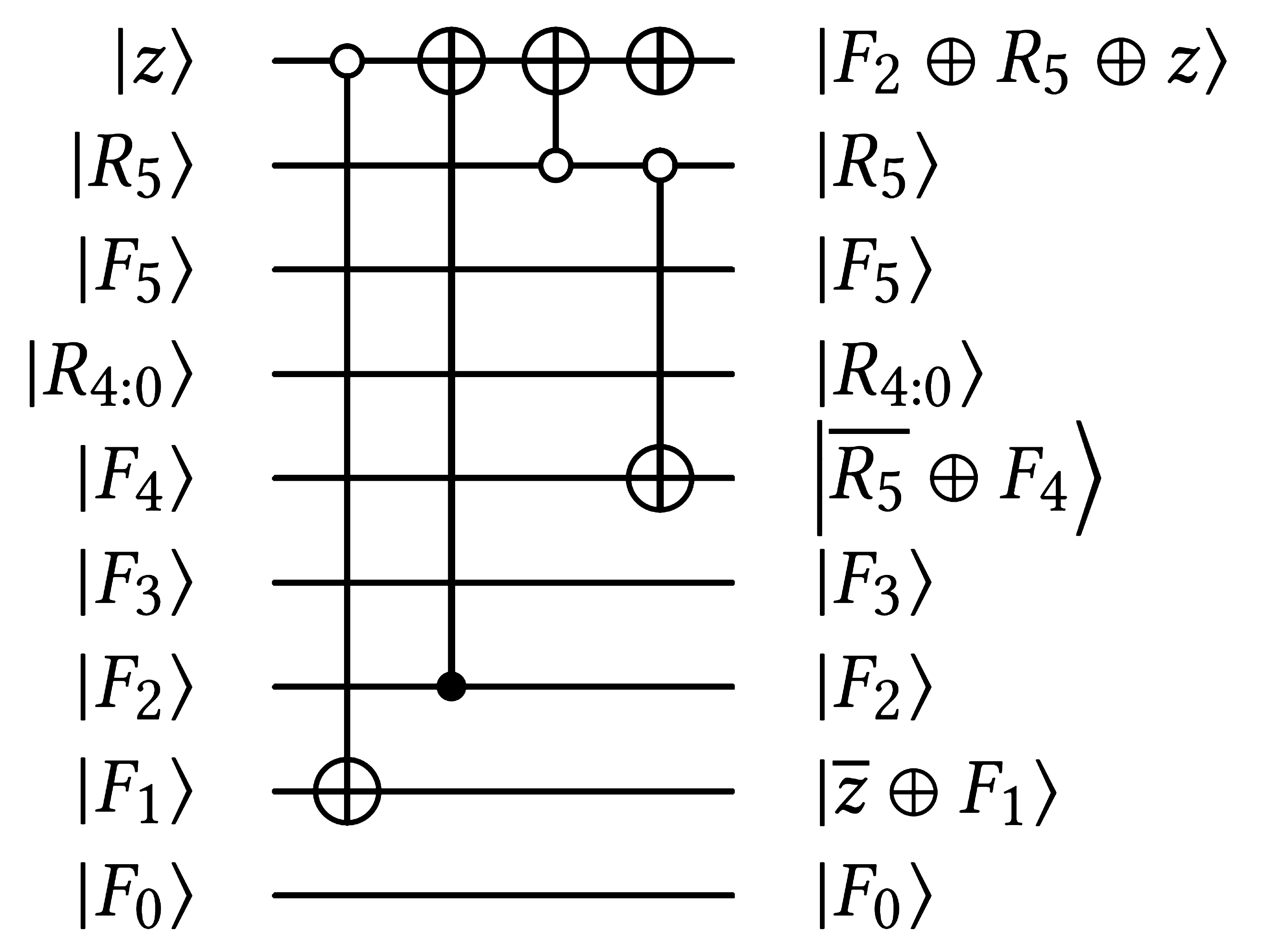}

\caption{After Steps 1 through 5}
\end{subfigure} \qquad \begin{subfigure}[th]{2.1in}

\includegraphics[width=2.27in]{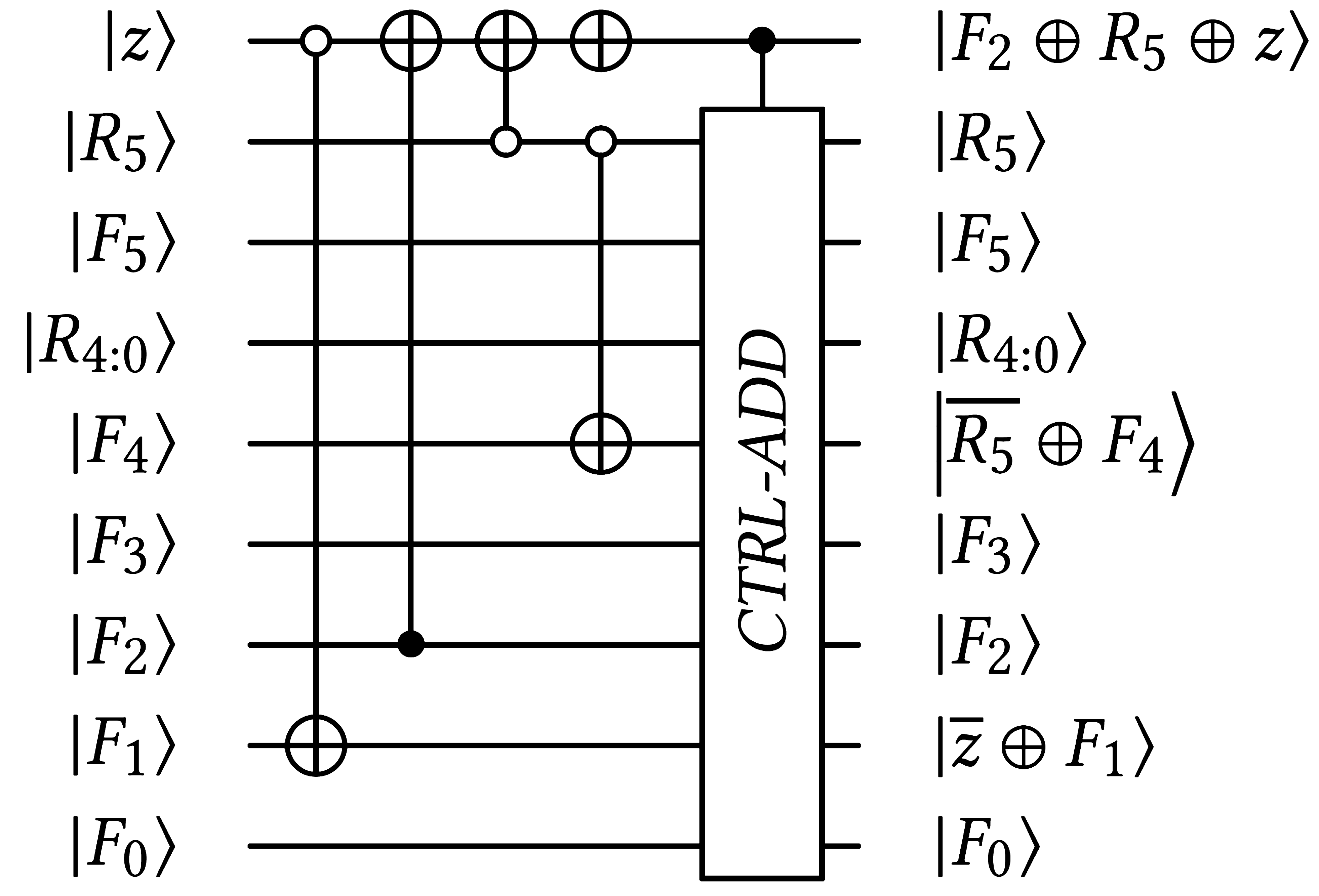}

\caption{After Step 6}
\end{subfigure}
\\
\begin{subfigure}[th]{4in}

\includegraphics[width=3.55in]{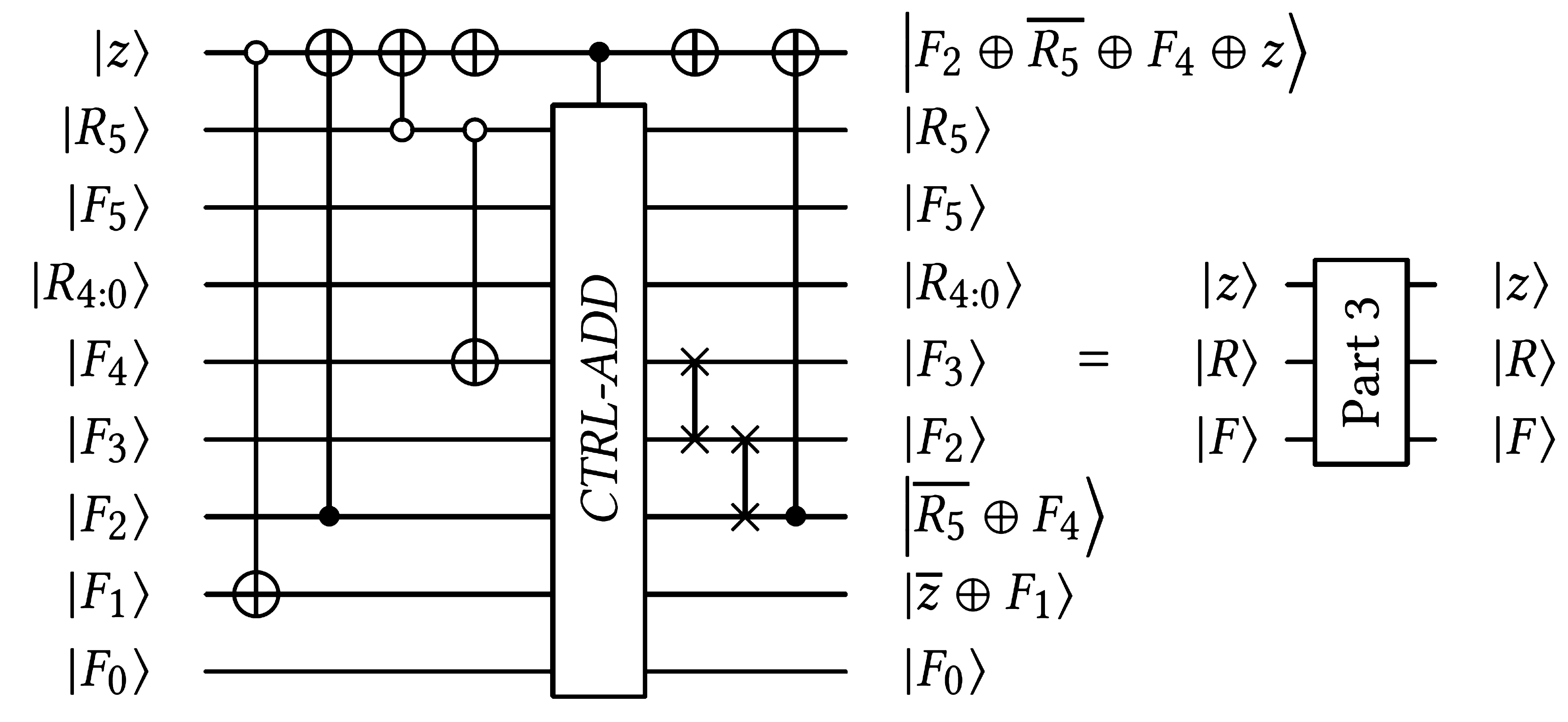}

\caption{After Steps 7 through 9. }
\end{subfigure}

\caption{Circuit generation of Part 3 of the proposed quantum square root circuit: Steps 1-9.  Quantum circuit and graphical representation are shown.  To keep Figures compact quantum register locations (such as $\ket{R_4}$ through $\ket{R_0}$) are represented as a single line and labeled accordingly (such as with $\ket{R_{4:0}}$) where possible.}
\label{sqrt-FIG4}
\end{figure}

This part only occurs once.  The quantum circuit for Part 3 takes quantum registers $\ket{R}$, $\ket{F}$ and $\ket{z}$ as inputs.  Part 3 has nine steps. Figure \ref{sqrt-FIG4} illustrates the generation of Part 3 with an example of a 6 bit square root circuit.

\begin{itemize}

\item Step 1: At locations $\ket{z}$ and $\ket{F_1}$ apply an inverted control CNOT gate such that location $\ket{z}$ is unchanged while location $\ket{F_1}$ now has the value $\ket{\overline{z} \oplus F_1}$.  This step restores $\ket{F_1}$ to its initial value such that $\ket{F}$ has the value: $ \ket{0} \cdots \ket{0} \ket{Y_{\frac{n}{2}-1}} \cdots \ket{Y_{1}}  \ket{0} \ket{1}$.  Thus, this step partially completes line 26 of Algorithm 1 when $\ket{R} < 0$ or partially completes line 31 otherwise.

\item Step 2: At locations $\ket{F_2}$ and $\ket{z}$ apply a CNOT gate such that location $\ket{F_2}$ is unchanged while location $\ket{z}$ now has the value $\ket{F_2 \oplus z}$  which simplifies to the value $0$. Step 1 and Step 2 prepare $\ket{z}$ and $\ket{F}$ for the IF statement in Algorithm 1.

\item Step 3: At locations $\ket{R_{n-1}}$ and $\ket{z}$ apply an inverted control CNOT gate such that location $\ket{F_2}$ is unchanged while location $\ket{z}$ now has the value $\ket{\overline{R_{n-1}} \oplus F_2 \oplus z}$ (where $\overline{R_{n-1}} \oplus F_2 \oplus z \equiv Y_{0}$).  This step prepares register $\ket{z}$ for use in subsequent steps.  

\item Step 4: At locations $\ket{R_{n-1}}$ and $\ket{F_{\frac{n}{2}+1}}$ apply an inverted control CNOT gate such that location $\ket{R_{n-1}}$ is unchanged while location $\ket{F_{\frac{n}{2}+1}}$ now has the value $\ket{\overline{R_{n-1}} \oplus F_{\frac{n}{2}+1}}$ (where $\overline{R_{n-1}} \oplus F_{\frac{n}{2}+1} \equiv Y_{0}$).  If $\ket{R_{n-1}} = 1$, this step continues execution of line 26 of Algorithm 1 and if  $\ket{R_{n-1}} = 0$, this step continues execution of line 31 of Algorithm 1.  Quantum register $\ket{F}$ will have the value: $\ket{0} \cdots \ket{0}  \ket{Y_{0}} \ket{Y_{\frac{n}{2}-1}} \cdots \ket{Y_{1}} \ket{0} \ket{1}$.  

\item Step 5: At location apply $\ket{z}$ apply a quantum NOT gate.  The value of $\ket{z}$ is now $\ket{R_{n-1} \oplus F_2 \oplus z}$ (where $R_{n-1} \oplus F_2 \oplus z \equiv \overline{Y_{0}}$).  This step prepares $\ket{z}$ for subsequent computations.  

\item Step 6: This step has the following two sub-steps.  

\begin{itemize}

\item Step 1: Apply quantum registers $\ket{F}$ and $\ket{R}$ to a quantum \textit{CTRL-ADD} circuit such that $\ket{F}$ is unchanged while $\ket{R}$ will hold the result of computation.

\item Step 2: At location $\ket{z}$ apply a quantum conditional addition (\textit{CTRL-ADD}) circuit such that the operation of the quantum \textit{CTRL-ADD} circuit is conditioned on the value at location $\ket{z}$.

\end{itemize}

After this step, if $\ket{R_{n-1}} = 1$, the quantum register $\ket{R}$ will equal $\ket{R} + \ket{F}$ (line 28 of Algorithm 1).  If $\ket{R_{n-1}} = 0$, the value in quantum register $\ket{R}$ is unchanged.  After this step, $\ket{R}$ will contain the remainder from calculating $Y$ (or $\sqrt{a}$).   

\item Step 7: At location $\ket{z}$ apply a quantum NOT gate.  The value of $\ket{z}$ is restored to the value $\ket{\overline{R_{n-1}} \oplus F_2 \oplus z}$ (where $\overline{R_{n-1}} \oplus F_2 \oplus z \equiv Y_{0}$).

\item Step 8: For $j = \frac{n}{2}+1 \text{ to } 3$:

At locations $\ket{F_{j}}$ and $\ket{F_{j-1}}$ apply a quantum SWAP gate such the values at locations $\ket{F_{j}}$ and $\ket{F_{j-1}}$ switch locations.  This step reorders the positions of the bits of $Y$ in $\ket{F}$.  Now, quantum register $\ket{F}$ will have the value: $\ket{0} \cdots \ket{0}  \ket{Y_{\frac{n}{2}-1}} \cdots \ket{Y_{0}} \ket{0} \ket{1}$.  Thus, when $\ket{R_{n-1}} = 1$, this step completes the execution of line 26 of Algorithm 1.   When $\ket{R_{n-1}} = 0$, this step completes the execution of line 31 of Algorithm 1.   After this step, $\ket{F}$ will contain the final value of $Y$ (or $\sqrt{a}$).

\item Step 9: At locations $\ket{F_2}$ and $\ket{z}$ apply a CNOT gate such that location $\ket{F_2}$ is unchanged while location $\ket{z}$ now has the value $\ket{\overline{R_{n-1}} \oplus F_2 \oplus F_4 \oplus z}$ which simplifies to $\ket{0}$.  This step completes the restoration of $\ket{z}$ to its initial value ($0$). 

\end{itemize}

Thus, the square root of $a$ is at locations $\ket{F_{\frac{n}{2}+1}}$ through $\ket{F_{2}}$ of quantum register $\ket{F}$ and the remainder of calculating the square root of $a$ is at quantum register $\ket{R}$.  The quantum register $\ket{z}$ along with locations $\ket{F_{n-1}}$ through $\ket{F_{\frac{n}{2}+2}}$ and locations $\ket{F_{1}}$ through $\ket{F_{0}}$ of quantum register $\ket{F}$ are restored to their initial values.  Thus, the proposed design methodology generates a quantum square root circuit that correctly implements the non-restoring square root algorithm.

\section{Cost Analysis}
\label{sqrt-cost}

\subsection{T-count Cost}

The proposed design methodology reduces the T-count by incorporating T gate efficient implementations of quantum \textit{CTRL-ADD} circuits and quantum \textit{ADD/SUB} circuits.
Garbageless and T gate optimized quantum \textit{ADD/SUB} and \textit{CTRL-ADD} circuits in the literature such as the designs in \cite{Thapliyal2016addsub} \cite{Saeedi2} and \cite{edgard2017multiplier} can be used in our proposed quantum square root circuit.  The T-count of the proposed quantum square root circuit is illustrated shortly for each part of the proposed design.  

\subsubsection{Part 1: Initial Subtraction}
		
		\begin{itemize}
			
			\item Steps 1 through 5 do not require T-gates. 
			
			\item Step 6 requires $42$ T gates.   We use a quantum \textit{ADD/SUB} circuit of T-count $14 \cdot n - 14$ in this step (where $n=4$). 
			
		\end{itemize}
	
\subsubsection{Part 2: Conditional Addition or Subtraction}
	The steps in this part are repeated $\frac{n}{2}-2$ times.   We show the T-count for the $i$th iteration of Part 2 where $2 \leq i \leq \frac{n}{2}-1$ 
	
	\begin{itemize}
			
		\item The $i$th iteration of 1 through 6 do not require T-gates.
					
			\item The T-count for the $i$th iteration of Step 7 is $14 \cdot ( 2 \cdot (i + 1)) -14$ which simplifies to $28 \cdot i + 14$.   We use a quantum \textit{ADD/SUB} circuit of T-count $14 \cdot n - 14$ in this step (where $n=2 \cdot (i+1)$).

		\end{itemize}

\subsubsection{Part 3: Reminder Restoration}
		
		\begin{itemize}
			
			\item Steps 1 through 5 do not require T-gates.
			\item The T-count for Step 6 is $21 \cdot n -14$. We use a quantum \textit{CTRL-ADD} circuit of T-count $21 \cdot n - 14$ in this step.
			\item Steps 7 through 9 do not require T-gates.
		\end{itemize}

\subsubsection{Calculation of T-count} 	
		
		To calculate the total T-count we add the total T-count for each part of the design.  The total T-count for Part 1 is $42$ (or $14 \cdot n - 14$ where $n=4$).  The total T-count for Part 2 is given as $\sum_{i = 2}^{\frac{n}{2}-1} 28 \cdot i +14$ and the total T-count for Part 3 is given as $21 \cdot n -14$.  Combining the total T-count for each part of the proposed quantum square root circuit results in the following expression: 
		
		\begin{equation}
		\left( \sum_{i = 1}^{\frac{n}{2}-1} 28 \cdot i +14 \right) + 21 \cdot n - 14
		\label{sqrt-equation:1}
		\end{equation} 
		
		
		The expression for the T-count (expression \ref{sqrt-equation:1}) can be simplified into the following expression:
		
		\begin{equation}
		\frac{7}{2} \cdot n^2 + 21 \cdot n - 28
		\label{sqrt-equation:2}
		\end{equation} 

\subsection{T-depth Cost}

We now calculate the T-depth for our proposed design.  Our proposed design is based on T-depth efficient designs of quantum \textit{ADD/SUB} circuits and quantum \textit{CTRL-ADD} circuits.  We determined that garbageless and T gate optimized quantum \textit{ADD/SUB} circuits in the literature such as the design in \cite{Thapliyal2016addsub} have a T-depth that is constant and independent of the circuit size $n$.  Thus, these \textit{ADD/SUB} circuits have T-depth of order $\mathcal{O}(1)$.  We determined as well that \textit{CTRL-ADD} circuits in the literature such as the design in \cite{edgard2017multiplier} scale as a function of circuit size $n$.  Thus, these \textit{CTRL-ADD} circuits have a T-depth of order $\mathcal{O}(n)$.  The T-depth of the proposed quantum square root circuit is illustrated shortly for each part of the proposed design.


\subsubsection{Part 1: Initial Subtraction}
		
		\begin{itemize}
			
			\item Steps 1 through 5 do not require T-gates. 
			
			\item Step 6 has a constant T-depth of $10$.  This T-depth is seen by locations $\ket{R_{n-2}}$ and $\ket{R_{n-3}}$ of quantum register $\ket{R}$.  We use a quantum \textit{ADD/SUB} circuit in this step.  The \textit{ADD/SUB} circuit has a constant T-depth $10$ that is independent of the circuit's size. 
			
		\end{itemize}
	
\subsubsection{Part 2: Conditional Addition or Subtraction}
	The steps in this part are repeated $\frac{n}{2}-2$ times.   We show the T-count for the $i$th iteration of Part 2 where $2 \leq i \leq \frac{n}{2}-1$ 
	
	\begin{itemize}
			
		\item The $i$th iteration of 1 through 6 do not require T-gates.
					
			\item Step 7 has a constant T-depth of $10$.  This T-depth is seen by locations $\ket{R_{n-2}}$ through $\ket{R_{n-2 \cdot i -1}}$ of quantum register $\ket{R}$.  We use a quantum \textit{ADD/SUB} circuit in this step.  The \textit{ADD/SUB} circuit has a constant T-depth $10$ that is independent of the circuit's size. 
			
			

		\end{itemize}

\subsubsection{Part 3: Reminder Restoration}
		
		\begin{itemize}
			
			\item Steps 1 through 5 do not require T-gates.
			\item Step 6 has a T-depth of $2 \cdot n$.  This T-depth is seen by quantum register $\ket{z}$.  We use a quantum \textit{CTRL-ADD} circuit of T-depth $2 \cdot n$ in this step.
			\item Steps 7 through 9 do not require T-gates.
		\end{itemize}

\subsubsection{Calculation of T-depth} 

We now illustrates the steps we use to determine the total T-depth for the proposed quantum square root circuit:

\begin{itemize}

\item Step 1: Calculate the T-depth for Part 1.  Part 1 has a T-depth of $10$.  This T-depth is seen by locations $\ket{R_{n-2}}$ and $\ket{R_{n-3}}$ of quantum register $\ket{R}$.

\item Step 2: Calculate the T-depth for Part 2.  Part 2 has a T-depth of $10 \cdot \left( \frac{n}{2}-2 \right)$ because Part 2 requires $\frac{n}{2}-2$ quantum \textit{ADD/SUB} circuits.  The total T-depth $10 \cdot \left( \frac{n}{2}-2 \right)$ simplifies to $5 \cdot n - 20$.  This T-depth is seen by locations $\ket{R_{n-2}}$ and $\ket{R_{n-3}}$ of quantum register $\ket{R}$.

\item Step 3: Calculate the T-depth for Part 3.  Part 3 has a T-depth of $2 \cdot n$.  This T-depth is seen by quantum register $\ket{z}$.

\item Step 4: Determine which qubits see the most T gate layers.  We find after comparing all the qubits in our proposed design quantum register $\ket{z}$ and quantum register locations $\ket{R_{n-2}}$ and $\ket{R_{n-3}}$ of $\ket{R}$ see the most T gate layers.

\item Step 5: Determine the total number of T gate layers seen by quantum register $\ket{z}$ in the proposed design.  Quantum register $\ket{z}$ will see a total of $2 \cdot n$ T gate layers because in Part 1 and Part 2, no T gates operate on quantum register $\ket{z}$.

\item Step 6: Determine the total number of T gate layers seen by quantum register locations $\ket{R_{n-2}}$ and $\ket{R_{n-3}}$ of $\ket{R}$ in the proposed design.  Quantum register locations $\ket{R_{n-2}}$ and $\ket{R_{n-3}}$ will see a total of $10$ T gate layers from Part 1, $5 \cdot n - 20$ T gate layers from Part 2 and $13$ T gate layers from Part 3.  The total T-depth for locations $\ket{R_{n-2}}$ and $\ket{R_{n-3}}$ is $5 \cdot n + 3$.  Quantum \textit{CTRL-ADD} circuits in the literature such as the design in \cite{edgard2017multiplier} present a constant T-depth to locations $\ket{R_{n-2}}$ and $\ket{R_{n-3}}$ of quantum register $\ket{R}$ when $\ket{R}$ is supplied as an input.  We use a quantum \textit{CTRL-ADD} circuit that presents a constant T-depth of $13$ to locations $\ket{R_{n-2}}$ and $\ket{R_{n-3}}$.

\item Step 7: Determine which qubits see the most T gate layers.  We determined that locations $\ket{R_{n-2}}$ and $\ket{R_{n-3}}$ see more T gate layers than register $\ket{z}$ because $5 \cdot n+3 > 2 \cdot n$.  The number of T gate layers on qubits with the most T gate layers will determine the T-depth for the proposed quantum square root circuit.

\end{itemize}    

Thus, our proposed design has a T-depth of $5 \cdot n + 3$ and this T-depth is seen by locations $\ket{R_{n-2}}$ and $\ket{R_{n-3}}$ of quantum register $\ket{R}$.

\begin{table}[tbhp]
\caption{Comparison of quantum square root circuits}
\label{sqrt-table:1}
\centering
\resizebox{\textwidth}{!}{
\begin{tabular}{lcccccc}
\midrule
design	&		&		T-count		&	& 	T-depth 		&	&	qubits		\\	\cmidrule{1-1} \cmidrule{3-7}
1	&		&	$	7 \cdot n^2 + 14 \cdot n	$	& 	&	$3 \cdot n + 8$		&	& 	$	\frac{1}{4} \cdot n^2 + 6 \cdot n - 2	$	\\	
2	&		&	$	420 \cdot n^2 + 168 \cdot n -364	$	& 	&		NA		&	& 	$	\approx 42 \cdot n + 10	$	\\[2pt]	
3	&		&	$	\frac{21}{4} \cdot n^2 + \frac{105}{2} \cdot n -42	$	&	&		NA		&	& 	$	\approx \frac{1}{2} n^2 + 7 \cdot n + 2	$	\\[2pt]	
4	&		&	$	\frac{21}{4} \cdot n^2 + \frac{7}{2} \cdot n -14	$	&	&		NA		&	& 	$	\approx \frac{1}{2} n^2 + 3 \cdot n + 4	$	\\[2pt]	
proposed	&		&	$	\frac{7}{2} \cdot n^2 + 21 \cdot n  - 28	$	&	&		$5 \cdot n + 3$		&	& 	$	2 \cdot n + 1	$	\\[2pt]	\bottomrule

\multicolumn{5}{l}{1 is the design by Sultana et al. \cite{Sultana}}\\ 
\multicolumn{5}{l}{2 is the design by Bhaskar et al. \cite{bhaskar}}\\
\multicolumn{5}{l}{3 is the first design by AnanthaLakshmi et al. \cite{ANANTHALAKSHM}}\\
\multicolumn{5}{l}{4 is the second design by AnanthaLakshmi et al. \cite{ANANTHALAKSHM}}\\
\multicolumn{5}{l}{Table entries are marked NA where a closed-form expression}		\\
\multicolumn{5}{l}{is not available for the T-depth.}		\\

\end{tabular}
}
\end{table}

\subsection{Cost Comparison}

		The comparison of the proposed quantum square root circuit with the current state of the art is illustrated in Table \ref{sqrt-table:1}.  To compare our proposed square root circuit against the existing designs by Sultana et al.\cite{Sultana} and AnanthaLakshmi et al. \cite{ANANTHALAKSHM}, we implemented the designs with Clifford+T gates.  We also apply the Bennett's garbage removal scheme (see \cite{Bennett1973trashremoval}) to remove the garbage output from the designs by Sultana et al. and AnanthaLakshmi et al.  The total qubit cost for each design by AnanthaLakshmi et al. are calculated by summing the garbage output produced by the controlled subtraction circuits and the circuit outputs.		 

To compare our proposed square root circuit against the existing design by Bhaskar et al.\cite{bhaskar}, we implemented the design with Clifford+T gates.  The square root design in Bhaskar et al. requires $5 \cdot \lceil log_2(b) \rceil$ multiplications and $3 \cdot \lceil log_2(b) \rceil$ additions (where $b$ is the number of bits of accuracy of the solution).  We use an implementation that has the lowest possible accuracy and thus let $b = 4$.  This is because the T gate and qubit costs increases as a function of solution accuracy.  Thus, the square root circuit based on the design by Bhaskar et al. requires $10$ multiplications and $6$ additions.  Bhaskar et al. did not specify a quantum adder or multiplier design for use in their square root quantum circuit design.  Therefore, to have a fair comparison against our work we use the quantum adder presented in \cite{Boss2} and the quantum multiplier shown in \cite{edgard2017multiplier}.  The quantum adder has a T-count of $14 \cdot n -7$, a qubit cost of $2 \cdot n + 1$ and produces no garbage output.  Further, the quantum multiplier has a T-count of $21 \cdot n^2 - 14$, a qubit cost of $4 \cdot n + 1$ and produces no garbage output.  We assume that given two inputs on quantum registers $\ket{a}$ and $\ket{b}$, the quantum multiplier will produce the product of $\ket{a}$ and $\ket{b}$ on $2 \cdot n +1$ ancillae.  The inputs $\ket{a}$ and $\ket{b}$ will maintain the same value at the end of computation.  Consequently, at the end of computation, the square root circuit based on the design by Bhaskar et al. will have garbage outputs.  We apply the Bennett's garbage removal scheme to remove the garbage output from the quantum circuit implementation of the design by Bhaskar et al.		
		
\subsubsection{Cost Comparison in Terms of T-count}		 

The T-count cost of the proposed quantum square root circuit and the designs by Sultana et al., Bhaskar et al. and AnanthaLakshmi et al. are of order $\mathcal{O}(n^2)$.  We compared the T-count cost of our proposed design methodology to the designs presented by Sultana et al., Bhaskar et al. and AnanthaLakshmi et al. for values of $n$ ranging from $4$ to $512$.  We calculated that our design achieves improvement ratios ranging from $33.33 \% $ to $49.61 \%$, $98.41 \%$ to $99.16 \% $, $33.84 \%$ to $55.56 \%$ and $0.00 \%$ to $32.64 \%$ compared to the designs by Sultana et al., Bhaskar et al. and AnanthaLakshmi et al.  

\subsubsection{Cost Comparison in Terms of Qubits}	

Table \ref{sqrt-table:1} also shows that our proposed design and the design by Bhaskar et al. have a qubit cost of order $\mathcal{O}(n)$ while the qubit cost for the designs by Sultana et al. and AnanthaLakshmi et al. are of order $\mathcal{O}(n^2)$.  We also compared the qubit cost of our proposed design methodology to the designs presented by Sultana et al., Bhaskar et al. and AnanthaLakshmi et al. for values of $n$ ranging from $4$ to $512$.  We calculated that our proposed design methodology achieves improvement ratios ranging from $65.38 \% $ to $98.51 \%$, $94.94 \%$ to $95.24 \% $, $76.32 \%$ to $99.24 \%$ and $62.50 \%$ to $99.23 \%$ compared to the designs by Sultana et al., Bhaskar et al. and AnanthaLakshmi et al. 

\subsubsection{Cost Comparison in Terms of T-depth}		 
		 
		 Table \ref{sqrt-table:1} illustrates how T-depth and qubit cost are linked and that minimizing one will result in an increase in the other resource cost measure.  Table \ref{sqrt-table:1} shows that the T-depth of our proposed design and the design by Sultana et al. are of order $\mathcal{O}(n)$.   Table \ref{sqrt-table:1} illustrates the trade-off between the T-depth and number of qubits.  However, the design by Sultana et al. is only able to achieve a constant factor of T-depth improvement against the proposed work at the expense of having a qubit cost of order $\mathcal{O}(n^2)$.  Our proposed design achieves a qubit cost of order $\mathcal{O}(n)$.  Thus, we significantly reduced the number of qubits in our proposed circuit and maintained a T-depth of the same order ($\mathcal{O}(n)$) as the work by Sultana et al.  
		 


\section{Conclusion}
\label{sqrt-ender}

In this work, we present a new design of a quantum square root circuit.  The proposed design has zero overhead in terms of garbage output.  The proposed design also requires fewer T gates and less qubits then the current state of the art.  The proposed quantum square root circuit has been formally verified.  The proposed quantum square root circuit could form a crucial component in the quantum hardware implementations of scientific algorithms where qubits and T-count are of primary concern.

\begin{acks}

The authors would like to thank the reviewers for their proposed suggestions that helped in further saving of T gates.

\end{acks}


\bibliographystyle{ACM-Reference-Format}
\bibliography{MULTbiblio.bib}

\appendix
\section{Example of the Non-Restoring Square Root Algorithm}
\label{sqrt-explain}

In this section, we present an example of Algorithm 1.  We shall illustrate the calculation of the square root of $26$.  We represent $26$ as a $6$ bit positive binary number in $2's$ complement ($a = 011010$).  The square root of $26$ is $5$ with a remainder of $1$.  At the end of computation, $R$ will have the remainder ($1$) from calculating the square root of $26$ and bit positions $F_4$ through $F_2$ of $F$ will contain the square root value $Y$ (where $Y = 5$).   

\begin{tabular}{cccl}
 & & & \\ \toprule
R & F & & Operations \\ \midrule
000001 & 000001 & & Assign $R = 0^4 a_{5} a_{4}$ and $F = 0^4 01$.  Where $0^4$ are $4$ zeros. \\
 & & & $a_5$ is the most significant bit of $a$. \\
000000 & 000001 & & Calculate $R = R - F$ \\
000010 & 000101 & & $i = 2$ and $R \geq 0$ so $Y_2 = 1$ \\
 & & & Assign $R =  00 R_{1} R_{0} a_{3} a_{2} $ and  $F = 0^3 Y_2 01$.  Where $0^3$ are $3$ zeros.  \\
 & & & Locations $R_{1} R_{0}$ in $R$ are shifted and reused. \\
111101 & 000101 & & Calculate $R = R - F$\\
110110 & 001011 & & $i = 1$ and $R < 0$ so $Y_1 = 0$ \\
 & & & Assign $R = R_{3} R_{2} R_{1} R_{0} a_{1} a_{0} $ and  $F = 00 Y_2 Y_1 11$ \\
 & & & Locations $R_{3}$ through $R_{0}$ in $R$ are shifted and reused. \\
000001 & 001011 & &Calculate $R = R + F$ \\
000001 & 010101 & &$R \geq 0$ so $Y_0 = 1$ \\
& & &Assign $F = 0 Y_2 Y_1 Y_0 01$ \\
000001 & 010101 & & Return $F$ and $R$  \\ \bottomrule
 & & \\

\end{tabular}

As expected $R = 000001$ which is the binary representation of the remainder $1$. Bit positions $F_4$ through $F_2$ of $F$ contain the values $101$ which is the binary representation of the number $5$ (the calculated square root value of $26$).   

\end{document}